\documentclass[american,english,aip,floatfix,jcp,reprint,longbibliography]{revtex4-1}
\usepackage[T1]{fontenc}
\usepackage[latin9]{inputenc}
\usepackage{units}
\usepackage{amsmath}
\usepackage{amssymb}
\usepackage{graphicx}

\makeatletter
\usepackage{graphicx}          
\usepackage{amsmath}    
\usepackage{amssymb}    
\usepackage{hyperref}   
\usepackage[capitalise]{cleveref}   

\usepackage{enumitem}
\usepackage{amsthm}
\usepackage{float}
\usepackage[capitalise]{cleveref}
\usepackage{appendix}
\usepackage{color}
\usepackage{soul}
\usepackage{bm}
\usepackage{textgreek}

\makeatother

\usepackage{babel}
\begin{document}
\title{Progress in deep Markov State Modeling: Coarse graining and experimental
data restraints}
\author{Andreas Mardt}
\affiliation{Department of Mathematics and Computer Science, Freie Universität
Berlin, Berlin, Germany}
\author{Frank Noé}
\affiliation{Department of Mathematics and Computer Science, Freie Universität
Berlin, Berlin, Germany}
\affiliation{Department of Physics, Freie Universität Berlin, Berlin, Germany}
\affiliation{Department of Chemistry, Rice University, Houston TX, 77005, United
States}
\email[correspondence to ]{frank.noe@fu-berlin.de}

\selectlanguage{english}%
\begin{abstract}
Recent advances in deep learning frameworks have established valuable
tools for analyzing the long-timescale behavior of complex systems
such as proteins. Especially the inclusion of physical constraints,
e.g. time-reversibility, was a crucial step to make the methods applicable
to biophysical systems. Furthermore, we advance the method by incorporating
experimental observables into the model estimation showing that biases
in simulation data can be compensated for. We further develop a new
neural network layer in order to build an hierarchical model allowing
for different level of details to be studied. Finally, we propose
an attention mechanism which highlights important residues for the
classification into different states. We demonstrate the new methodology
on an ultralong molecular dynamics simulation of the Villin headpiece
miniprotein.\foreignlanguage{american}{}
\end{abstract}
\maketitle

\section{Introduction}

The thermodynamics and kinetics of large biological macromolecules
can be studied in full spatiotemporal detail with molecular dynamics
(MD) simulations \citep{ShirtsPande_Science2000_FoldingAtHome,Phllips_JCC05_NAMD,HarveyDeFabritiis_JCTC09_ACEMD,BuchEtAl_JCIM10_GPUgrid,Shaw_Science10_Anton,EastmanEtAl_JCTC13_OpenMM,SalomonFerrerWalker_JCTC13_Amber14,PronkEtAl_Bioinf13_Gromacs4.5,DoerrEtAl_JCTC16_HTMD}.
The combined progress in high-throughput MD simulations and in analysis
frameworks such as Markov state models (MSMs) \citep{schuette:j-comput-phys:1999:conformational-dynamics,swope:jpcb:2004:markov-model-theory,noe:jcp:2007:markov-models,ChoderaEtAl_JCP07,Bowman_JCP09_Villin,PrinzKellerNoe_PCCP11_Perspective,SchuetteEtAl_JCP11_Milestoning,BowmanEnsignPande_JCTC2010_AdaptiveSampling,Pande_Methods10_MSMs,WeberFackeldeySchuette_JCP17_SetfreeMSM,ferguson2011nonlinear,razavi2014computational},
Master-equation models \citep{ChekmarevIshidaLevi_JPCB04_MasterEq,SriramanKevrekidisHummer_JPCB109_6479,buchete-hummer:2008:coarse-master-equations}
and closely related approaches \citep{NoeEtAl_PMMHMM_JCP13,WuNoe_MMS14_TRAM1,RostaHummer_DHAM,WuNoe_JCP15_GMTM,WuEtAL_PNAS16_TRAM,BowmanEnsignPande_JCTC2010_AdaptiveSampling,EVandenEijnden_MMS04_Metastability,TiwaryParrinello_PRL14_MetadynamicsDynamics,RibeiroTiwary_JCP18_RAVE,chen2018molecular}
has led to the successful characterizations of the kinetics of folding
proteins \citep{noe:pnas:2009:ww-domain,Bowman_JCP09_Villin,LindorffLarsenEtAl_Science11_AntonFolding},
protein-ligand association \citep{BuchFabritiis_PNAS11_Binding,SilvaHuang_PlosCB_LaoBinding,PlattnerNoe_NatComm15_TrypsinPlasticity,Tiwari_PNAS14_KineticsProteinLigandUnbinding},
and even protein-protein association \citep{PlattnerEtAl_NatChem17_BarBar}.
These methods benefit from the facts that they do not require a rigorous
definition of reaction coordinates \citep{SarichNoeSchuette_MMS09_MSMerror}
and allow the extraction of experimental observables which can be
connected to structural changes of the system \citep{NoeEtAl_PNAS11_Fingerprints,KellerPrinzNoe_ChemPhysReview11,Zhuang_JPCB11_MSM-IR,LindnerEtAl_JCP13_NeutronScatteringI,ChoderaNoe_JCP09_MSMstatisticsII,voelz2010unfolded,zhou2017bridging}.

Additionally, the MSM approach can handle non-equilibrium data by
approximating the transition density $p_{\tau}(\mathbf{x},\mathbf{y})$
of a Markov process:
\begin{equation}
p_{\tau}(\mathbf{x},\mathbf{y})=\mathbb{P}(\mathbf{x}_{t+\tau}=\mathbf{y}\mid\mathbf{x}_{t}=\mathbf{x}),\label{eq:cond_prob}
\end{equation}

that models the conditional probability of the system transitioning
to configuration $\mathbf{y}$ when starting in configuration $\mathbf{x}$
after a time interval $\tau$ , called lag-time. 

Whereas the construction of MSMs has previously been done manually
via a challenging and potentially error prone pipeline of feature
selection, dimension reduction, clustering, estimating a transition
matrix $\mathbf{P}$, etc, it can be largely automated using variational
approaches that optimize MSMs to best resolve the rare event processes
\citep{SchwantesPande_JCTC13_TICA,PerezEtAl_JCP13_TICA}. Firstly,
the variational approach for conformation dynamics (VAC) has been
developed to find optimal model parameters \citep{NoeNueske_MMS13_VariationalApproach,NueskeEtAl_JCTC14_Variational}.
It was generalized by the variational approach for Markov processes
(VAMP) which increases the scope to non-reversible and non-stationary
dynamics \citep{Wu2019}.

An end-to-end deep learning method called VAMPnets was proposed which
replaces the MSM building through training a neural network mapping
the configurations $\mathbf{x}$ to a low dimensional state space
$\boldsymbol{\chi}(\mathbf{x})$ \citep{MardtEtAl_VAMPnets}. This
state space resembles the state space of an ordinary MSM by having
a softmax output function which results in fuzzy state assignments
of each configuration, which can be interpreted as state probabilities
$\mathbf{p}(t)=\boldsymbol{\chi}(\mathbf{x}_{t})$. A VAMPnet can
be trained by maximizing the VAMP score to find an optimal state space
allowing the linear propagation of the state probabilities via a transition
matrix:
\begin{equation}
\mathbf{p}^{T}(t+\tau)=\mathbf{p}^{T}(t)\mathbf{P},\label{eq:Koopman_eq}
\end{equation}
which gives access to the kinetics of the system by studying the properties
of $\mathbf{P}$. However, the matrix is not guaranteed to satisfy
stochastic properties, i.e. it can have negative or entries larger
than $1$. Further advances have been proposed, which do not aim to
replace the whole pipeline, but instead finding optimal features to
build an ordinary MSM via the rest of the pipeline \citep{Chen_2019},
or which propose transferable feature functions across chemical space
\citep{xie2019graph}. Recently, a method was introduced which allows
VAMPnets with physical constraints to be constructed, addressing two
major issues:
\begin{enumerate}
\item The transition matrix $\mathbf{P}$ can be enforced to have exclusively
non-negative entries thus being a stochastic matrix. Therefore, further
analysis using transition path theory is possible \citep{EVandenEijnden_TPT_JStatPhys06,MetznerSchuetteVandenEijnden_TPT,NoeSchuetteReichWeikl_PNAS09_TPT}.
\item Time-reversibility (detailed balance) can be enforced in the model
when the underlying dynamics obeys microscopic reversibility, but
the data is sampled out of equilibrium. 
\end{enumerate}
The method allows to select either of the two constraints independently,
which enables the user to build four different classes of models.
In this work we will focus on the reversible deep MSM (revDMSM), which
has been already successfully employed to a disordered protein\citep{lohr2021kinetic},
to applying both constraints and thereby allowing the application
of transition path theory to study rates of interesting processes.

However, in the field of MSMs for protein analysis further developments
have been made which need to be transfered to the case of a deep learning
framework. Here, we address two major limitations of VAMPnets and
deep MSMs that have been addressed for classical MSMs and are important
ingredients in MSM applications.

On the one hand, several methods have been proposed to coarse-grain
an existing MSM using spectral clustering \citep{Deuflhard_LinAlgAppl_PCCA,DeuflhardWeber_LAA05_PCCA+,FackeldeyWeber_WIAS17_GenPCCA},
which allows the user to study the system on different degrees of
detail. Here, we develop a new layer structure for neural networks
to coarse-grain the existing model by still obeying the constraints
and keeping it trainable end-to-end.

On the other hand, it has been proposed and showed that incorporating
existing experimental information into the model estimation helps
to overcome systematic errors in the force field \citep{OlssonEtAl_PNAS17_AugmentedMarkovModels,HummerKoefinger_JCP15_Bayesian,pitera2012use,boomsma2014combining,beauchamp2014bayesian,cavalli2013molecular,leung2016rigorous,PrinzKellerNoe_PCCP11_Perspective}.
Consequently, we show how different classes of observables can be
included into the training routine of a revDMSM.

Furthermore, the black box nature of deep learning frameworks fueled
the development of methods aiming at understanding individual decisions
of the neural network \citep{montavon2018methods,ribeiro2016should,lundberg2017unified,fong2017interpretable,mnih2014recurrent,xu2015show,bau2017network,kindermans2017learning,selvaraju2017grad,zhang2019interpreting}.
These fall into two categories, the post hoc analysis of the network
and the simultaneously trained attention mechanism. Here, we focus
on the later to optimize it for the application on proteins in the
context of a revDMSM. 

In summary, we will advance the revDMSM by the following contributions
based on a study of a small protein:
\begin{enumerate}
\item Comparison between building a MSM via a revDMSM and a VAMPnet.
\item Analyzing the slow processes. 
\item Applying transition path theory to estimate folding and unfolding
rates.
\item Deriving an algorithm in order to built an hierarchical model to allow
an easier interpretation of the resulting model.
\item Proposing an attention mechanism which supports the understanding
of how important each residue is for the dynamic classification.
\item Developing an algorithm to incorporate experimental observations into
the model estimation.
\end{enumerate}

\section{Methods}

\begin{figure}
\includegraphics[width=0.8\columnwidth]{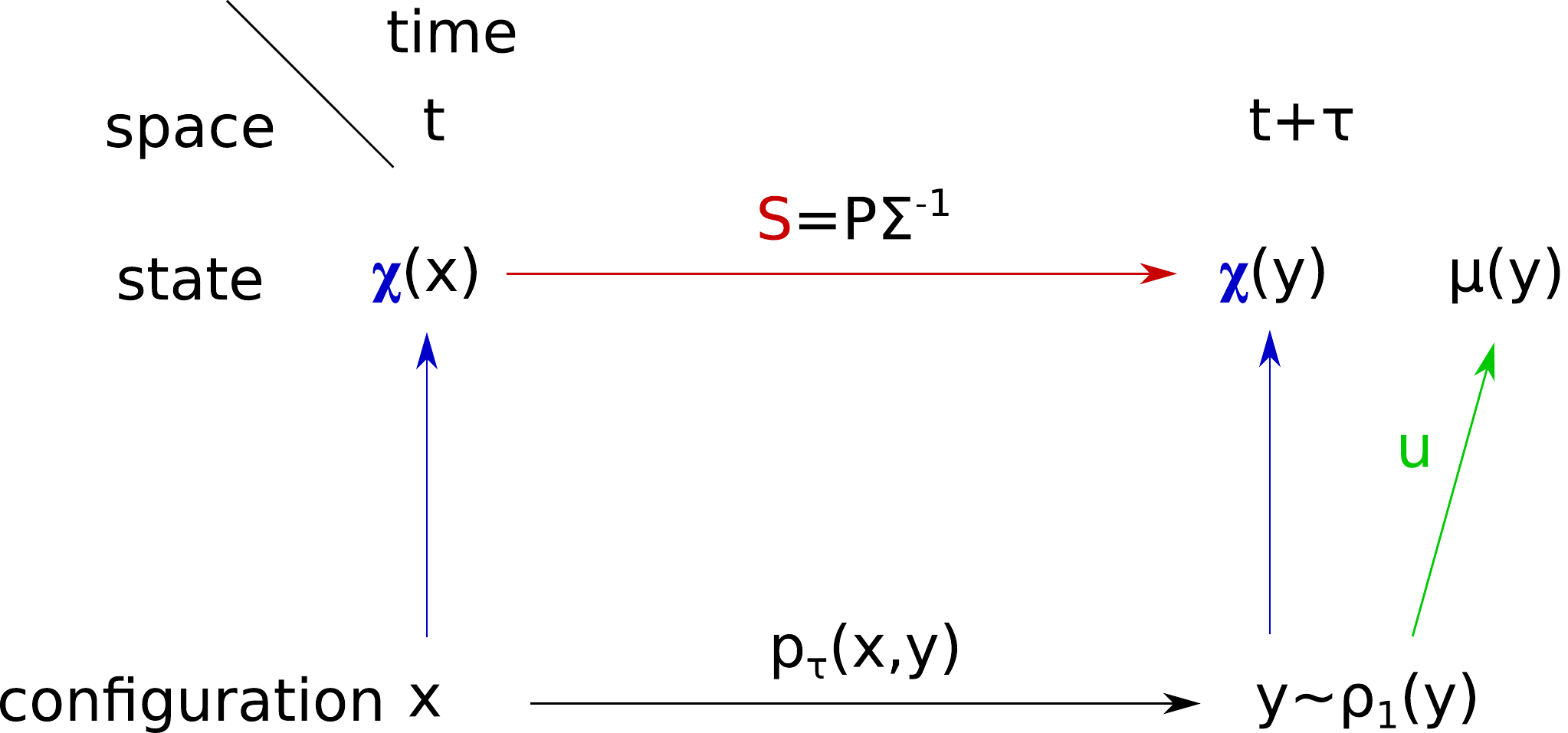}

\caption{Schematic of the probability model given by a revDMSM. The conditional
probability distribution $p_{\tau}(\mathbf{x},\mathbf{y})$ of jumping
to configuration $\mathbf{y}$ after a time interval $\tau$ given
a system in configuration $\mathbf{x}$ is approximated by transferring
the estimation into a trainable state space $\boldsymbol{\chi}(\mathbf{x})$.
The starting configuration $\mathbf{x}$ is mapped to the state space
$\boldsymbol{\chi}(\mathbf{x})$ and consecutively propagated by the
transition matrix $\mathbf{P}$. The probability of the propagated
state vector belonging to a configuration $\mathbf{y}$ is given by
the scalar product with the state vector $\boldsymbol{\chi}(\mathbf{y})$
weighted by the stationary distribution $\mu(\mathbf{y})$. The stationary
distribution is given via the trainable vector $\mathbf{u}$ and the
transition matrix via the trainable matrix $\mathbf{S}=\mathbf{P}\boldsymbol{\Sigma}^{-1}$,
where $\boldsymbol{\Sigma}=\int\boldsymbol{\chi}(\mathbf{y})\mu(\mathbf{y})\boldsymbol{\chi}(\mathbf{y})^{T}\mathrm{d}\mathbf{y}$
normalizes the probability distribution, i.e. integrating the distribution
over all possible configurations $\mathbf{y}$ evaluates to $1$.\label{fig:model}}
\end{figure}

As proposed in Ref. \citenum{mardt2020deep} the transition density
of a Markov process is approximated by:
\begin{equation}
p_{\tau}(\mathbf{x},\mathbf{y})=\boldsymbol{\chi}(\mathbf{x})^{T}\mathbf{S}\boldsymbol{\chi}(\mathbf{y})\boldsymbol{\chi}(\mathbf{y})^{T}\mathbf{u}\rho_{1}(\mathbf{y}),\label{eq:model}
\end{equation}

which models the probability of the system transitioning to configuration
$\mathbf{y}$ when starting in configuration $\mathbf{x}$ after a
time interval $\tau$, called lag-time. The function $\boldsymbol{\chi}(\mathbf{x})$
represented by a neural network maps a configuration to a fuzzy state
space of dimension $m$ similar to a VAMPnet \citep{MardtEtAl_VAMPnets}.
The trainable vector $\mathbf{u}\in\mathbb{R}^{m}$ reweights the
empirical distribution $\rho_{1}(\mathbf{y})$ to the learned stationary
distribution:
\begin{equation}
\mu(\mathbf{y})=\boldsymbol{\chi}(\mathbf{y})^{T}\mathbf{u}\rho_{1}(\mathbf{y}).\label{eq:stat}
\end{equation}

The trainable matrix $\mathbf{S}\in\mathbb{R}^{m\times m}$ gives
access to the transition matrix $\mathbf{P}$ between the states:
\begin{align}
\mathbf{P} & =\mathbf{S\boldsymbol{\Sigma}},\label{eq:Koopman_matrix_calculation}
\end{align}
\begin{align}
\boldsymbol{\Sigma} & =\int\boldsymbol{\chi}(\mathbf{y})\mu(\mathbf{y})\boldsymbol{\chi}(\mathbf{y})^{T}\mathrm{d}\mathbf{y}\nonumber \\
 & =\int\boldsymbol{\chi}(\mathbf{y})\rho_{1}(\mathbf{y})\boldsymbol{\chi}(\mathbf{y})^{T}\mathbf{u}\boldsymbol{\chi}(\mathbf{y})^{T}\mathrm{d}\mathbf{y}\label{eq:equil_cov}
\end{align}
being the equilibrium covariance matrix of $\boldsymbol{\chi}(\mathbf{y})$
(Fig. \ref{fig:model}). 

\subsection{Reversible deep Markov State Models}

Since $\boldsymbol{\chi}(\mathbf{x})$ should be a fuzzy state assignment
and the stationary distribution and the transition matrix have to
be normalized, the following constraints have to be fulfilled\citep{mardt2020deep}:
\begin{enumerate}
\item Normalized state vector: $\boldsymbol{\chi}(\mathbf{x})^{T}\mathbf{1}=1.$
\item Normalized stationary distribution: $\bar{\boldsymbol{\chi}}^{T}\mathbf{u}=1$,
where $\bar{\boldsymbol{\chi}}=\mathbb{E}\left[\boldsymbol{\chi}(\mathbf{x}_{t+\tau})\right]$
is the empirical state probability, which results in $\int\mu(\mathbf{y})\mathrm{d}\mathbf{y}=\bar{\boldsymbol{\chi}}^{T}\mathbf{u}=1$.
\item Normalized transition matrix: $\mathbf{S}\mathbf{C}_{\tau\tau}^{\prime}\mathbf{u}=\mathbf{1}$
where $\mathbf{C}_{\tau\tau}^{\prime}=\mathbb{E}\left[\boldsymbol{\chi}(\mathbf{x}_{t+\tau})\boldsymbol{\chi}(\mathbf{x}_{t+\tau})^{\top}\right]$
is the empirical covariance matrix of $\boldsymbol{\chi}(\mathbf{x}_{t+\tau})$.
As a result the transition matrix preserves probability mass by means
of $\mathbf{P1}=\mathbf{\mathbf{S}\mathbf{C}_{\tau\tau}^{\prime}\mathbf{u}=\mathbf{1}}$.
\end{enumerate}
The general approach allows the training of different classes of models.
Here, we will focus on the revDMSM, where the matrix $\mathbf{S}$
has to be additionally symmetric $\mathbf{S}=\mathbf{S}^{T}$ and
$\mathbf{S},\mathbf{u}$, and $\boldsymbol{\chi}$ have to be non-negative.
The constraints will be matched by a proper choice of architectures
for the different trainable parts.

Two different losses for training are introduced: the maximum likelihood
(ML), where the probability of observing the data according to the
model is maximized, and the VAMP-E score, which is used throughout
the paper. The training procedure of such a model includes the pretraining
of a VAMPnet, which we have slightly modified (Appendix \ref{par:Pretraining-the-VAMPnet}).

It has been proven that the proposed model is an universal approximator
for reversible Markov processes\citep{mardt2020deep}, which makes
it therefore a promising candidate to study biological systems. Furthermore,
it was demonstrated that it yields asymptotically unbiased results
even in the case of many short trajectories allowing the analysis
of parallel simulated data.

\subsection{Incorporating experimental observables in the model estimation}

For an experiment such as fluorescence, chemical shift in NMR, IR
spectroscopy the conformational dynamics are mapped onto an observable
$a$. In the following we assume that $a$ has a scalar value associated
with conformation $\mathbf{x}$, $a(\mathbf{x})$. The generalization
to vector- or tensor-valued observables is straightforward. In equilibrium
the ensemble average will be measured by the experiment, which can
be approximated by a weighted sum with the equilibrium weights $\mu(\mathbf{x}_{t})$
for simulation data with $\sum_{t=1}^{T}\mu(\mathbf{x}_{t})=1$:
\begin{equation}
\mathbb{E}[a]\approx\sum_{t=1}^{T}\mu(\mathbf{x}_{t})a(\mathbf{x}_{t}).\label{eq:Exp_value}
\end{equation}

Since simulations carry a systematic bias due to the force field approximation
and finite sampling, the estimated value from a simulation may differ
from the experiment. However, if experimental information for some
observables is available, they can be incorporated into the model
estimation. By possibly removing the bias from the simulation calculations
of observables not included in the estimation procedure could be improved.
Therefore, we propose to extent the loss function by the objective
to match the observables measured by experiments:
\begin{equation}
L_{\text{total}}=L_{\text{MSM}}+\sum_{i}\lambda_{i}||O_{i}-\mathbb{E}[a_{i}]||^{2},\label{eq:exp_loss}
\end{equation}
where $O_{i}$ is the ensemble average measured in experiment for
the $i$th observable, where each observable can be weighted by $\lambda_{i}$
which encodes uncertainty about the observation. $L_{\text{MSM}}$
is either the VAMP-E or ML loss as proposed in the original paper. 

Depending on where the bias is to be expected, the loss can be used
to train for all instances $\boldsymbol{\chi},\mathbf{u},\mathbf{S}$.
Otherwise, if e.g. the state definition seems trustworthy and a reweighting
should be enough to counteract the bias $\mathbf{u}$ could be trained
by the additional loss keeping $\boldsymbol{\chi}$ and $\mathbf{S}$
fixed. 

The same approach can be used for available kinetic information through
time-correlation experiments. The expectation value of these time-correlations
can be expressed via the simulation and the model as:
\begin{align}
\mathbb{E}[a(t)a(t+k\tau)] & \approx\sum_{t_{1}=1}^{T}\mu(\mathbf{x}_{t_{1}})a(\mathbf{x}_{t_{1}})\sum_{t_{2}=1}^{T}p_{k\tau}(\mathbf{x}_{t_{1}},\mathbf{x}_{t_{2}})a(\mathbf{x}_{t_{2}})\label{eq:autocorr}\\
 & =\mathbf{a}^{T}\mathbf{X}^{k}(\tau)\mathbf{a},\nonumber 
\end{align}

where $\mathbf{a}$ gives the average value of the observable within
each state and $\mathbf{X}^{k}(\tau)$ is the unconditional probability
to jump between the states \citep{PrinzKellerNoe_PCCP11_Perspective}.
We can estimate both quantities via our model as:
\begin{align}
\mathbf{a}_{i} & =\sum_{t}a(\mathbf{x}_{t})\frac{\boldsymbol{\chi}_{i}(\mathbf{x}_{t})\mu(\mathbf{x}_{t})}{\sum_{t'}\boldsymbol{\chi}_{i}(\mathbf{x}_{t'})\mu(\mathbf{x}_{t'})},\label{eq:auto_corr_update}\\
\mathbf{X}(\tau) & =\boldsymbol{\Sigma}_{t}\mathbf{P},\nonumber 
\end{align}

with $\boldsymbol{\Sigma}_{t}$ being the equilibrium covariance matrix
of $\boldsymbol{\chi}(\mathbf{x}_{t})$: $\boldsymbol{\Sigma}_{t}=\int\boldsymbol{\chi}(\mathbf{x}_{t})\mu(\mathbf{x}_{t})\boldsymbol{\chi}(\mathbf{x}_{t})^{T}\mathrm{d}\mathbf{x}_{t}$.
In the case of a normalized time correlation with $\bar{a}=a-\mathbb{E}[a]$
the term reads:
\begin{equation}
\frac{\mathbb{E}[\bar{a}(t)\bar{a}(t+\tau)]}{\mathbb{E}[\bar{a}(t)^{2}]}=\frac{\mathbf{\bar{a}}^{T}\mathbf{X}^{k}(\tau)\mathbf{\bar{a}}}{\sum_{t=1}^{T}\mu(\mathbf{x}_{t})\bar{a}^{2}(\mathbf{x}_{t})}.\label{eq:norm_auto_corr}
\end{equation}

Additionally, it is of great interest to incorporate experimentally
measurable relaxation timescales $t_{i}$ into the model which are
directly related to the eigenvalues $\lambda_{i}$ of the transition
matrix $\mathbf{P}$ via $t_{i}=-\log(\lambda_{i})/\tau$. However,
the transition matrix $\mathbf{P}$ is not Hermitian and therefore
estimating gradients of its eigenvalues is numerically unstable and
not supported by the main deep learning tools such as \textit{PyTorch}
or \textit{tensorflow}. Fortunately, from Eq. (\ref{eq:Koopman_matrix_calculation})
it follows:
\begin{equation}
\bar{\mathbf{S}}=\boldsymbol{\Sigma}^{1/2}\mathbf{S}\boldsymbol{\Sigma}^{1/2}=\boldsymbol{\Sigma}^{1/2}\mathbf{P}\boldsymbol{\Sigma}^{-1/2},\label{eq:hermitian_similar}
\end{equation}

which shows that $\mathbf{P}$ is similar to the Hermitian matrix
$\bar{\mathbf{S}}$ and therefore we can optimize for the eigenvalues
of $\bar{\mathbf{S}}$ instead. A connection between folding/unfolding
rates and timescales is given in Appendix \ref{par:Connection-between-timescales}.

\subsection{Coarse graining}

For interpreting and understanding the model it can be helpful to
build an hierarchical state splitting of the model, where the system
can first be studied on a coarse and subsequently on a finer level.
This structure can be obtained by training independent models with
different output sizes and comparing the resulting states \citep{MardtEtAl_VAMPnets}.
We propose instead to make use of the given loss and learn the coarse
graining on the fly. Given a model with $m$ output nodes we want
to learn a matrix $\mathbf{M}\in\mathbb{R}^{mxn}$, which maps the
output of the model $\boldsymbol{\chi}^{m}\in\mathbb{R}^{m}$ to a
coarser model with $n<m$ states via:
\begin{equation}
(\boldsymbol{\chi}^{n})^{T}=(\boldsymbol{\chi}^{m})^{T}\mathbf{M}.\label{eq:coarse-graining}
\end{equation}

The coarse graining matrix can be trained by using the proposed loss
functions or simply the VAMP-2 score in the case of a VAMPnet. Furthermore,
we want that $\mathbf{M}_{ij}>0$ and $\sum_{j}\mathbf{M}_{ij}=1$,
which is ensured by defining trainable weights $\mathbf{m}\in\mathbb{R}^{m\times n}$
and squeeze them through a softmax function:
\begin{equation}
\mathbf{M}_{ij}=\frac{\exp(m_{ij})}{\sum_{l}\exp(m_{il})}.\label{eq:cg_norm}
\end{equation}

However, when training a reversible model with the parameters $\mathbf{u}^{m}$
and $\mathbf{S}^{m}$ there is no need to retrain $\mathbf{u}^{n}$
and $\mathbf{S}^{n}$ for the finer model. Instead, it is preferable
to ensure consistency between the models regarding the stationary
and transition density. Starting with the stationary distribution
$\mu^{m}(\mathbf{x})\overset{!}{=}\mu^{n}(\mathbf{x})$, it follows:
\begin{align}
(\boldsymbol{\chi}^{m}(\mathbf{x}))^{T}\mathbf{u}^{m} & =(\boldsymbol{\chi}^{n}(\mathbf{x}))^{T}\mathbf{u}^{n}=(\boldsymbol{\chi}^{m}(\mathbf{x}))^{T}\mathbf{M}\mathbf{u}^{n}\label{eq:cg_u}\\
\Rightarrow\mathbf{u}^{m} & =\mathbf{Mu}^{n},\nonumber 
\end{align}

which can be solved via the pseudoinverse of $\mathbf{M}=\mathbf{UDV}^{T}$:
\begin{equation}
\mathbf{u}^{n}=\mathbf{VD}^{*T}\mathbf{U}^{T}\mathbf{u}^{m}=\mathbf{G}\mathbf{u}^{m},\label{eq:cg-solve-u}
\end{equation}
where $\mathbf{D}^{*}$ has the reciprocal value of every non-zero
element in $\mathbf{D}$.

A similar derivation can be done for $\mathbf{S}$ given that $p_{\tau}^{m}(\mathbf{x},\mathbf{y})\overset{!}{=}p_{\tau}^{n}(\mathbf{x},\mathbf{y})$
and $\mu^{m}(\mathbf{x})=\mu^{n}(\mathbf{x})$:
\begin{align}
(\boldsymbol{\chi}^{m}(\mathbf{x}))^{T}\mathbf{S}^{m}\boldsymbol{\chi}^{m}(\mathbf{y}) & =(\boldsymbol{\chi}^{n}(\mathbf{x}))^{T}\mathbf{S}^{n}\boldsymbol{\chi}^{n}(\mathbf{y})\label{eq:cg-S}\\
 & =(\boldsymbol{\chi}^{m}(\mathbf{x}))^{T}\mathbf{M}\mathbf{S}^{n}\mathbf{M}^{T}\boldsymbol{\chi}^{m}(\mathbf{y})\\
\Rightarrow\mathbf{S}^{m} & =\mathbf{MS}^{n}\mathbf{M}^{T}\nonumber \\
\mathbf{S}^{n} & =\mathbf{GS}^{m}\mathbf{G}^{T}.\nonumber 
\end{align}

However, both $\mathbf{u}^{n}$ and $\mathbf{S}^{n}$ still need to
fulfill the constraints. Since the pseudoinverse represents the least
square solution, the constraints could be violated. Therefore, we
propose to renormalize the previous results\citep{mardt2020deep}:
\begin{align}
\mathbf{u}^{n} & =\frac{\mathbf{u}^{n}}{(\bar{\boldsymbol{\chi}}^{n})^{T}\mathbf{u}^{n}}\nonumber \\
\mathbf{S}^{n} & =\mathbf{S}^{n}+\text{diag (\ensuremath{\mathbf{s}})}\label{eq:cg-renorm}\\
s_{i} & =\frac{1-\sum_{k}S_{ik}^{n}v_{k}^{n}}{v_{i}^{n}}.\nonumber 
\end{align}

In order to make the result more robust it is favorable to train $\mathbf{u}^{m},\mathbf{S}^{m},\mathbf{M}_{i}$
simultaneously for the sum of the VAMP-E scores of all models. However,
if it is preferable to retrain $\mathbf{u}^{n}$ and $\mathbf{S}^{n}$,
the consistency expressions should be taken into account in the loss
function.

Furthermore, we suggest to initiate the weights of $\mathbf{M}$ according
to the PCCA+ memberships \citep{RoeblitzWeber_AdvDataAnalClassif13_PCCA++}.
Thereby, we avoid the network becoming stuck in an unfavorable solution
during the beginning of training. 

\section{Results}

\subsection{Overview}

Below we demonstrate how a revDMSM can be applied and expanded by
our proposed methods to the Villin dataset provided by \citep{LindorffLarsenEtAl_Science11_AntonFolding}
(folded structure Fig. \ref{fig:correction} a). The input for $\boldsymbol{\chi}$
is $\exp(-\mathbf{d})$, where $\mathbf{d}$ is the minimal heavy
atom distance between all residues \citep{MardtEtAl_VAMPnets}. Firstly,
we will study the ability of the method to guarantee a reversible
MSM and compare the result against a VAMPnet in a low data regime.
Afterwards, we will build an hierarchical model and applying our proposed
attention mechanism to it. Based on this, we study folding and unfolding
rates via transition path theory. Finally, we will look at the effects
of including ground truth observables into the training. 

\subsection{Implementation}

The methods were implemented using \textit{PyTorch} \citep{paszke2019pytorch}.
For the full code and details about the neural network architecture,
hyperparameters and training routine, please refer to \href{https://github.com/markovmodel/deepmsm}{https://github.com/markovmodel/deepmsm}.
In general, we used the Adam optimizer \citep{adam}, a batchsize
of 10000, and a 6 layer deep neural network with a constant width
of 100 nodes with the ELU activation function for $\boldsymbol{\chi}$.
When using an attention network it has the same architecture as $\boldsymbol{\chi}$
except of the output size and a window size of 4 motivated by the
fact that the shortest beta strands and a helix turn is about four
residues long.

The whole simulation data is randomly split threefold into $70\%$
training, $20\%$ validation and $10\%$ test set data. The validation
data is used to tune hyperparameters and enact an early stopping mechanism.
The results are reported for the test set.

\subsection{Obtaining real eigenvalues and positive entries in the transition
matrix via a revDMSM}

\begin{figure*}
\includegraphics[width=0.9\textwidth]{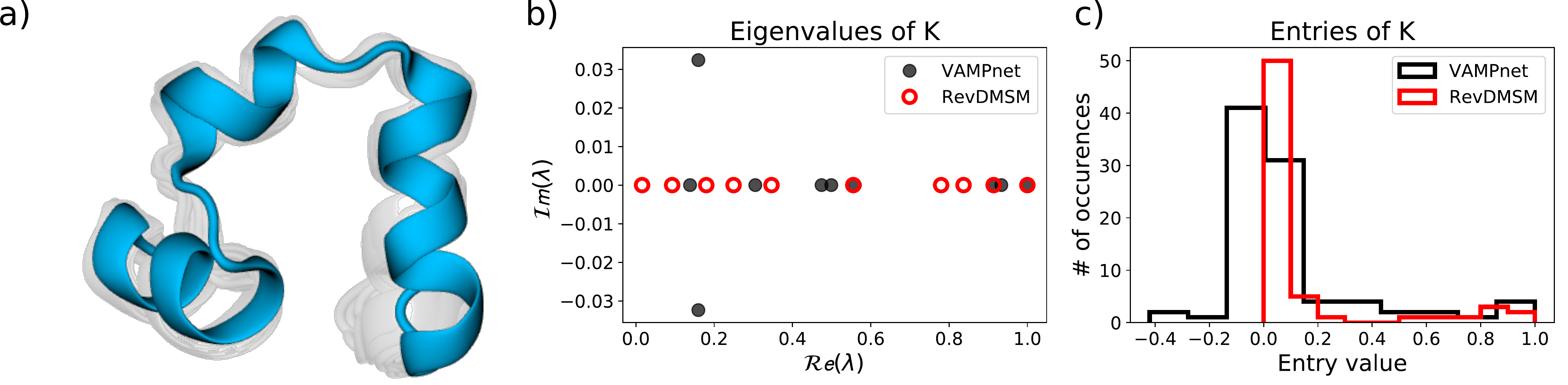}

\caption{Incorporating constraints on the matrix $\mathbf{P}$ to enforce a
reversible model in the regime of poorly sampled systems. a) Folded
structure of Villin b) Imaginary and real part of the spectrum and
c) entries of the matrix $\mathbf{P}$ of the non-reversible VAMPnet
and the revDMSM with a skip of 100 frames and 10 states (1 frame$=\tau=20\ \unit{ns}$).\label{fig:correction}}
\end{figure*}
To simulate an insufficient sampled example, we used a skip of 100
frames (1frame$=20\ \unit{ns}$) and 10 output nodes. We train a regular
VAMPnet and a revDMSM on the training data with an early stopping
given by the performance on the validation set and estimate the resulting
transition matrix on the test set at a time-lag of $\tau=20\ \unit{ns}$.
The eigenvalues of the transition matrix are always real for the revDMSM,
whereas in the case of a VAMPnet pairwise complex eigenvalues may
occur (\ref{fig:correction} b). Furthermore, the distribution of
the entries demonstrate how the revDMSM in contrast to the VAMPnet
guarantees values between $0$ and $1$, which can be interpreted
as probabilities (\ref{fig:correction} c). The insufficient sampling
leads in the case of the VAMPnet to a non-reversible model, where
the transition matrix is not a stochastic matrix. The revDMSM model
does not suffer from these shortcomings, nevertheless the constraints
imposed on the model result in slightly lower eigenvalues, which can
be expected since the constraints impair the ability to approximate
the eigenfunctions of the underlying operator.

\subsection{Building a MSM and testing its validity for a VAMPnet and a revDMSM}

\begin{figure*}
\includegraphics[width=0.9\textwidth]{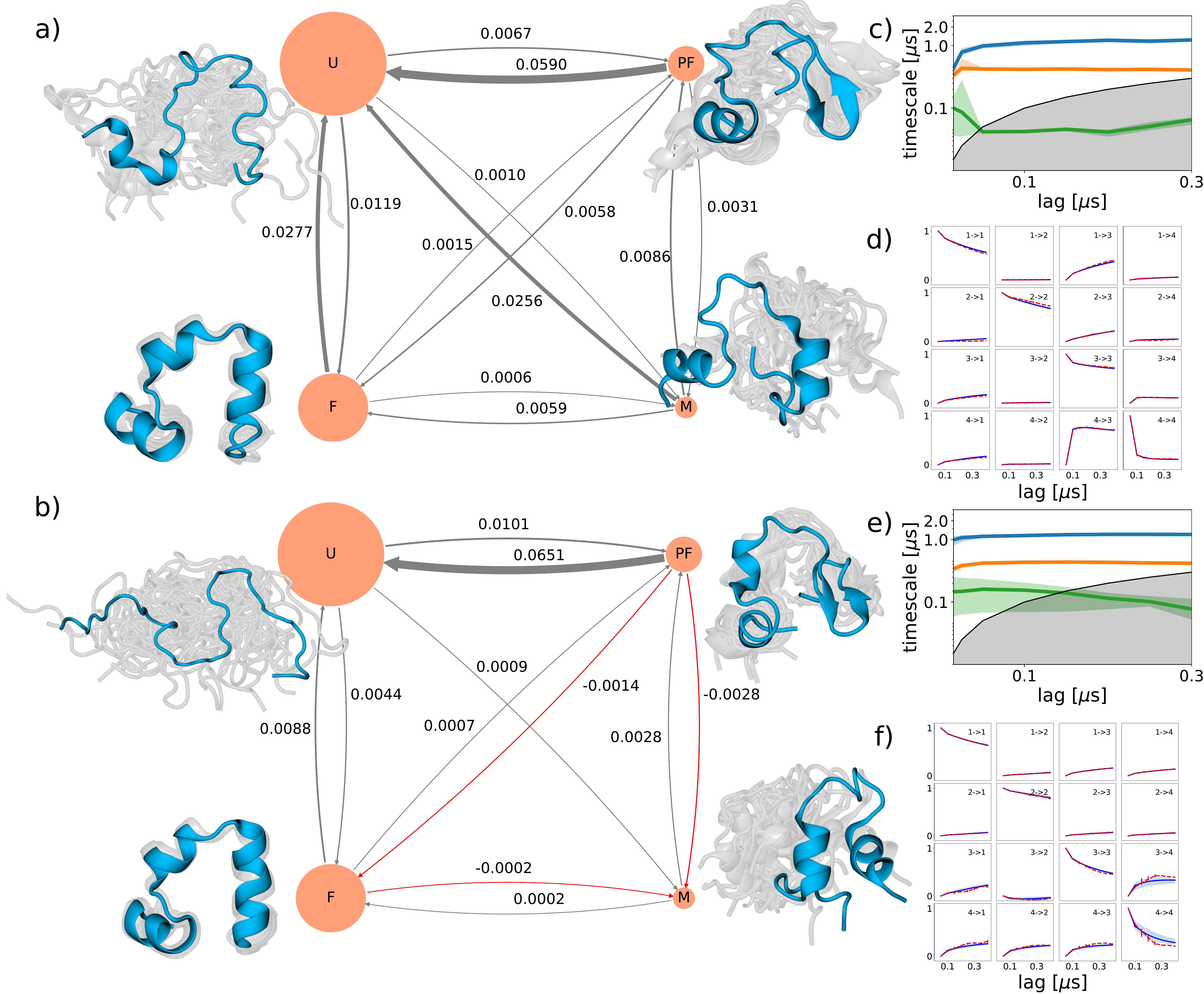}

\caption{Comparison of building a MSM with non-reversible VAMPnets and revDMSM
with a skip of 25 frames (1 frame$=5\ \unit{ns}$) and 4 states. State
network of the revDMSM model a) and VAMPnet b), the size of a state
corresponds to the stationary distribution, the arrows are the transition
rates of the $\mathbf{P}$ matrix (red arrows indicate negative entries).
Additionally, 10 representative structures aligned according to their
secondary structure are shown next to the states. Although trained
independently both feature functions identify a folded state (F),
and unfolded state (U), a partially folded state (PF), and a misfolded
state (M). Model validation through the implied timescales of the
c) revDMSM and e) VAMPnets model and the CK-test d) and f) at a base
model estimated at $\tau_{\text{MSM}}=50\ \unit{ns}$. Errors are
estimated over 10 runs.\label{fig:comp_msm}}
\end{figure*}
Based on the same data as above but with a skip of 25 frames (1 frame$=\tau=5\ \unit{ns}$),
we built a MSM with 4 output nodes using a revDMSM and a VAMPnet,
respectively. By inspecting the state network connected by their transition
probabilities we observe again negative transition probabilities for
the VAMPnet (red arrows Fig. \ref{fig:comp_msm} b). However, the
implied timescale and the CK test confirm the ability of both models
to predict the long-time kinetics. Furthermore, the models agree upon
the timescales within the 70th percentile estimated over 10 runs and
discover similar metastable states, where 10 aligned representative
structures are depicted next to each state (Fig. \ref{fig:comp_msm}
a, b). Although each model uses different trained feature functions
$\boldsymbol{\chi}$ they both identify a folded state (F), an unfolded
state (U), a partially folded state (PF), and a misfolded state (M)
characterized by a helix including the amino acids 20LEU and 21PRO
which form a coil in the folded state.

\subsection{Building an hierarchical model with an interpretable attention mechanism}

\begin{figure*}
\includegraphics[width=0.9\paperwidth]{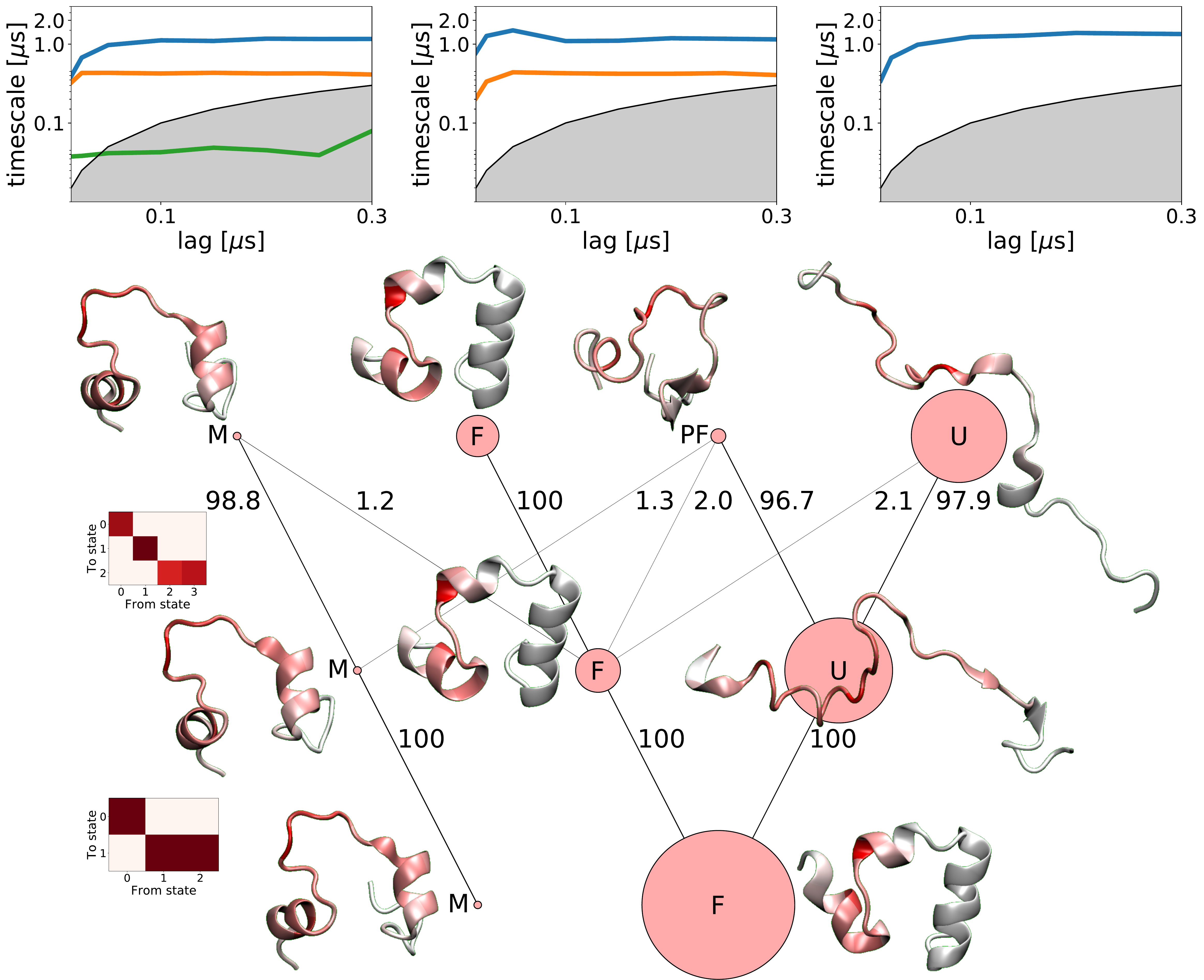}

\caption{Hierarchical model from 4 to 3 to 2 states with additional attention
analysis. The top row shows the implied timescales for each model
where the slowest timescales are conserved when coarse grained. Below,
the network graph is depicted where each node represents a state:
folded (F), unfolded (U), partially folded (PF), and misfolded (M).
The connections between the nodes represent how the states contribute
to the coarser representation, where the probability of belonging
to the coarse grained state is attached to it (probabilities $<1$\%
are omitted). Additionally, a graphical representation of the matrices
$\mathbf{M}$ are depicted beside it (white-low, red-high probability).
For each state the most likely configuration is shown, where the color
indicates the attention weight learned during training (white-low
red-high attention). \label{fig:Coarse-graining-from}}
\end{figure*}
In order to demonstrate the application of an hierarchical model we
coarse grain a 4 state revDMSM to a 3 state and consecutively to a
2 state model. Additionally, we incorporate an attention mechanism
into the architecture (Sec. \ref{subsec:Interpretability-via-attention}).
After training the 4 state model at $\tau=50\ \unit{ns}$ we train
the coarse-graining matrices with the VAMP-E score consecutively with
the pseudoinverse method. Finally, we simultaneously update $\mathbf{u}$
and $\mathbf{S}$ from the 4 state model and the two coarse graining
matrices to maximize the sum of the VAMP-E scores of all three models. 

For the estimation of the implied timescales of all models it is sufficient
to exclusively retrain $\mathbf{u}$ and $\mathbf{S}$ from the 4
state model to optimize the sum of the individual scores. The 3 and
2 state model conserve both the highest timescales as expected (Fig.
\ref{fig:Coarse-graining-from} top row). Thereby, the 2 state model,
where the unfolded state is missing, indicates that the slowest timescale
is not connected to the folding process. Instead, the state of the
unfolded structure appears in the 3 state model emphasizing the nature
of folding of the second timescale.

Below the timescales we depict a graphical representation of the hierarchical
model, where each node represents a state, where the same nomenclature
is used as above. The connections between the nodes represent the
coarse graining matrix, where the numbers encode the probability that
the state belongs to the coarse grained state (connections with a
probability less than $1$\% are not displayed). For reasons of clarity
we added additionally visual representations of the matrices beside
it, where the color encodes the probability (white-low, red-high).
There, it becomes evident that the coarse-grain matrices have a very
sparse structure. The large values seem to be in agreement with the
observation in the original paper that revDMSMs tend to rather hard
assignments of states. However, the hard assignment supports the interpretation
of an hierarchical model. 

Next to the nodes we added the structures of Villin with the highest
probability for that particular state. There, the hierarchical splitting
is visible: structure elements of the higher model hierarchy are preserved
in the lower level.

Furthermore, we depict the attention weight of each residue in the
color scheme of the structures (red-high, white-low). It is worth
mentioning, that the attention mechanism needs to highlight areas
which are important to distinguish all four metastable states. This
implies that the absence of a specific secondary structure could be
important and therefore highlighted. 

Remarkably, our attention mechanism detects 13ARG as important for
the folded structure (F), which we found aligns well with the folding
process (Sec. \ref{subsec:Augment-the-revDMSM}). The misfolded state
(M) shows high attention at the amino acids 20LEU and 21PRO being
part of a helix which form a coil in the folded state. The last residues
of the chain seem to be bad descriptors for the dynamics which seems
reasonable due to their more flexible nature.

In general, the network assigns high attention mainly to regions where
states have themselves secondary structure or where they lack the
structure other states have, e.g. the middle helix of the folded state.

\subsection{Approximation of the leading eigenfunctions via a revDMSM}

\begin{figure*}
\includegraphics[width=0.9\textwidth]{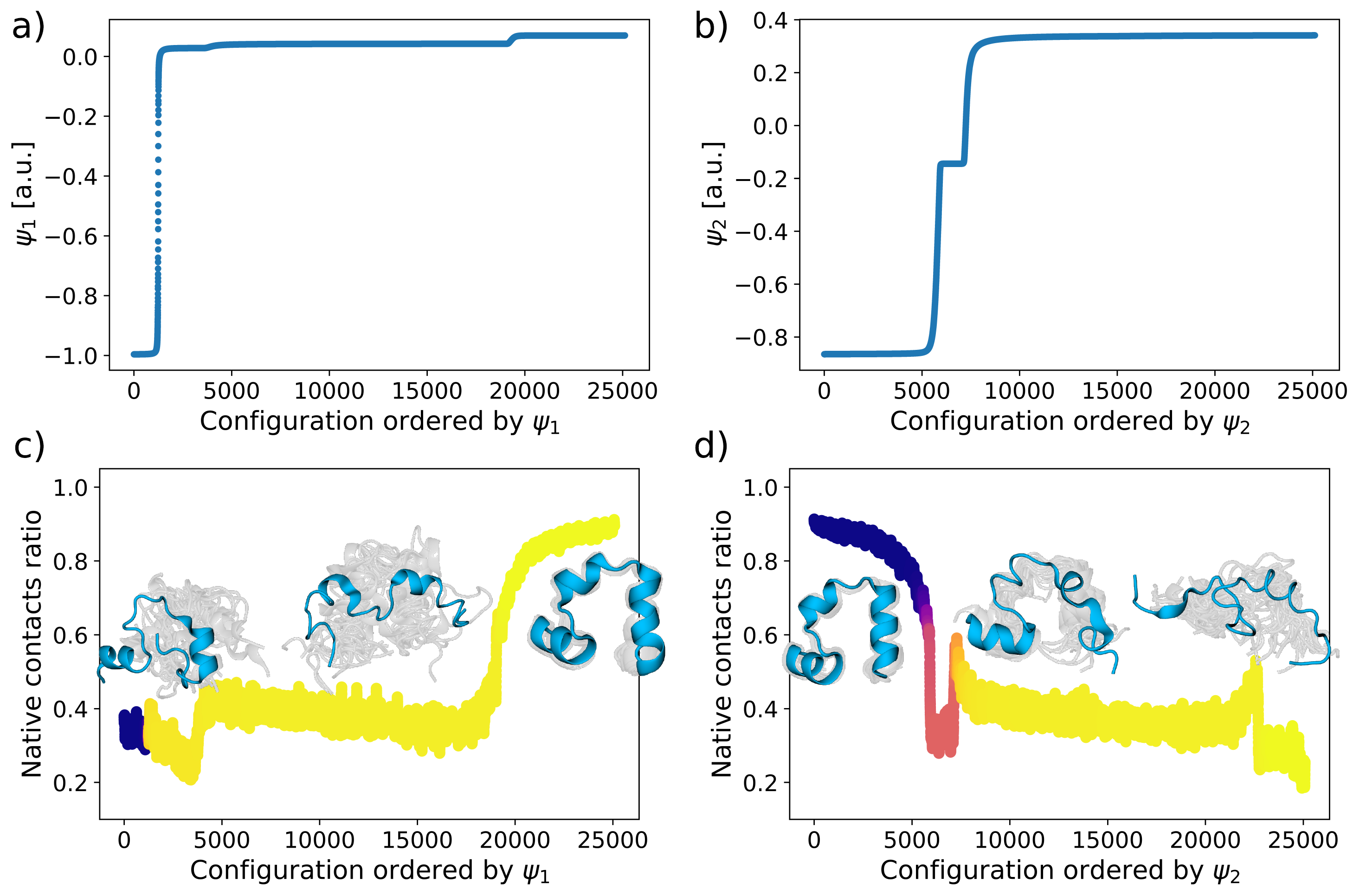}

\caption{Estimating the two slowest eigenfunctions with a revDMSM on the Villin
dataset with a skip of 25 frames ($\tau=50\ \unit{ns}$) and 4 states.
a), b) Eigenfunctions against frames ordered by their eigenfunction
value (fake trajectory along the process). c), d) The ratio of formed
native contacts with a mean window over 30 frames is plotted against
the same x-axis, where the color encodes the value of the eigenfunction.
Representative structures for three regions are shown.\label{fig:eigfuncs}}
\end{figure*}
Driven by the observation that the slowest process might not be the
folding process, we test the ability of the above revDMSM to approximate
the two slowest eigenfunctions of the 4 state model of Villin. In
order to visualize the process the frames are ordered by their eigenfunction
value (Fig. \ref{fig:eigfuncs} a, b) and plotted against the ratio
of how many native contacts are formed (Fig. \ref{fig:eigfuncs} c,
d). We define residues being in contact, which are at least two amino
acids apart in the chain and closer than $0.45\;\mathrm{nm}$ with
respect to their closest heavy atoms. In favor of a smoother result,
we apply a mean window over 30 frames. For both eigenfunctions 10
aligned representative structures for three regions along the process
are shown.

The results confirm that the slowest eigenfunction is not the folding
process \citep{husic2019deflation}. Instead, the analysis reveals
a process from the misfolded via the unfolded to the folded structure.
However, the second slowest process represents the expected folding
process from folded to unfolded.

\subsection{Estimation of folding rates\label{par:Estimation-of-folding}}

We can study via transition-path-theory (TPT) the folding and unfolding
rates of the system. We built a three state revDMSM, where $\boldsymbol{\chi}$
was pretrained at a time-lag of $\tau=5\ \unit{ns}$ and $\mathbf{u}$
and $\mathbf{S}$ at a lag-time of $\tau=100\ \unit{ns}$. The model
has to be constituted of at least three states because the folding
process is the second slowest timescale in the Villin trajectory.
Given the model we estimate from the transition matrix $\mathbf{P}$
the mean first passage time with the \textit{PyEMMA} package \citep{SchererEtAl_JCTC15_EMMA2}
for the folding and unfolding process:
\begin{align}
\tau_{\text{folding}} & =(3.0\pm0.3)\;\mu\text{s},\label{eq:res-mfpt}\\
\tau_{\text{unfolding}} & =(1.0\pm0.1)\;\mu\text{s,}\nonumber 
\end{align}

where the error is given via the standard deviations over 10 runs.
The original paper reports a folding rate of $(2.8\pm0.5)\;\mu\text{s}$
and unfolding rate of $(0.9\pm0.2)\;\mu\text{s}$ which is in perfect
agreement with our findings, although they utilize a handcrafted definition
of states and simply measure the average lifetime of the folded and
unfolded state and the transition time as the average of all events
in the trajectory \citep{LindorffLarsenEtAl_Science11_AntonFolding}.

\subsection{Estimating deep MSMs with experimental restraints\label{subsec:Augment-the-revDMSM}}

\begin{figure*}
\includegraphics[height=0.5\textheight]{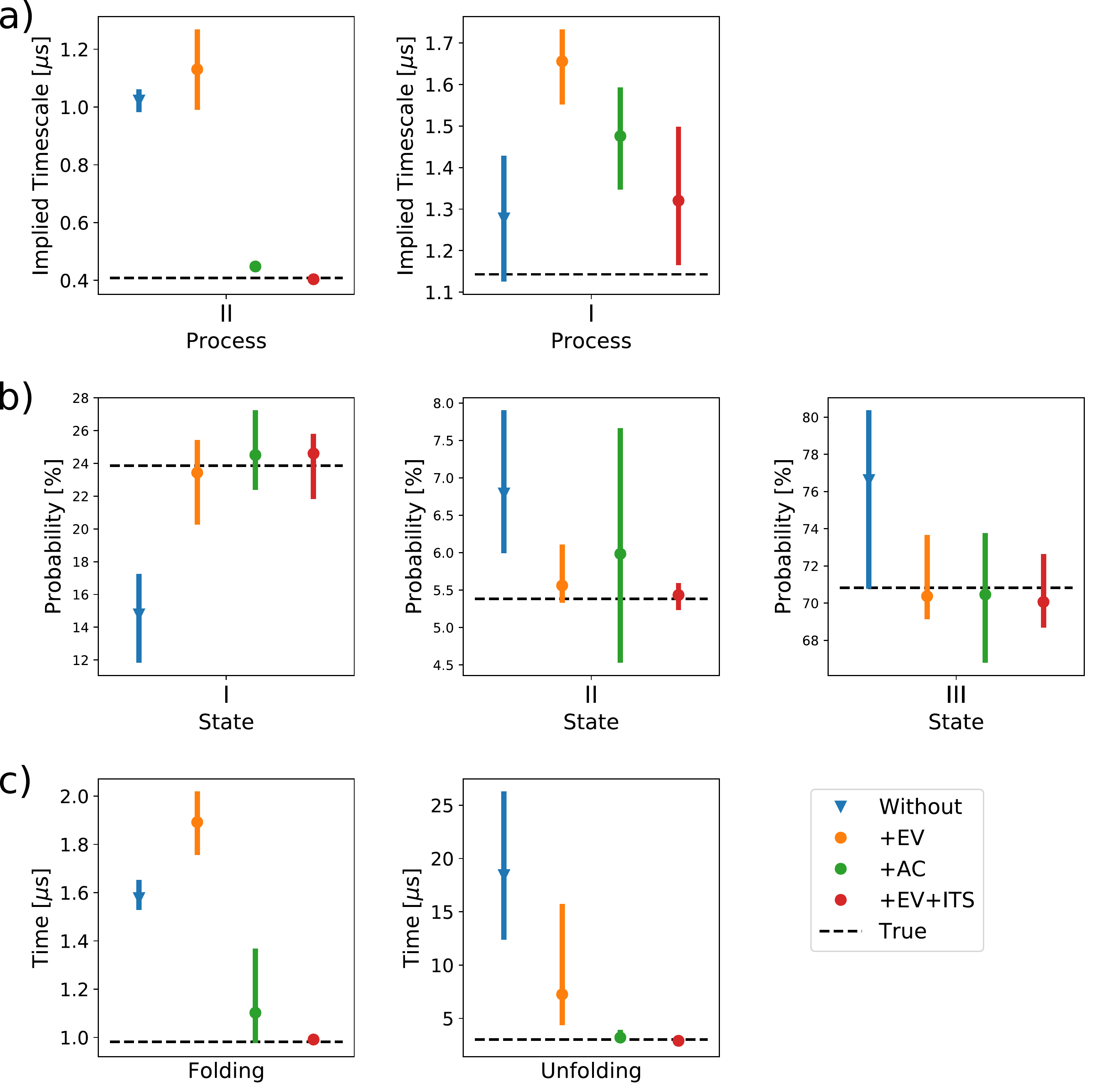}

\caption{Estimating observables with biased data and comparing estimates between
models incorporating some of them into the training routine (orange,
green, red) and a model incorporating none (blue). The data is biased
by removing folding/unfolding events from the trajectory to simulate
a probable higher transition barrier originating from possible erroneous
force fields. The true values (dashed black line), ideally from an
experiment, were estimated from a model trained on the whole unbiased
data set. a) two slowest timescales, where process II is the interesting
folding process, which was used for the red model as training observable.
Therefore, the better approximation of that process is expected. b)
state probabilities and therefore a stationary observable, which was
not used during training. However, using the expectation value of
the contact 1LEU-13ARG being formed improves the estimation of the
stationary properties (orange, red). c) folding and unfolding rates
estimated via the transition path theory not used for training, where
incorporating the timescale or the autocorrelations of the contact
1LEU-13ARG staying formed or unformed improves the estimation significantly
(red and green). The error bars indicate the 68th percentile.\label{fig:obs-stat}}
\end{figure*}
In order to mimic the situation of having ground truth values preferably
from an experiment but biased simulation data for model estimation,
we treat expectation values from the whole long simulation as ground
truth. However, we imitate the situation of a simulation which overestimates
transition energy barriers compared to the ground truth by removing
three fourth of folding and unfolding events from the full trajectory.
We detect these events by inspecting the sign changes of the second
slowest eigenfunction.

Since we perturbed by our data modification the folding/unfolding
process, we focus on observables related to it. Some of them serve
as additional information for training, the others as a validation
set to compare the performance on them:
\begin{enumerate}
\item The kinetics via the implied timescales (ITS) of the two slowest processes.
\item The stationary distribution of three predefined states.
\item The folding and unfolding rates via TPT analysis.
\item The expectation value (EV) of contact 1LEU-13ARG being formed.
\item The autocorrelation (AC) of the contact 1LEU-13ARG staying formed
or unformed.
\end{enumerate}
The contact 1LEU-13ARG is chosen because it correlates well with the
folding eigenfunction and could be possibly experimentally observed
by attaching fluorescence labels and conducting an fluorescence correlation
experiment. A contact is said to be formed if the minimal residue
distance is shorter than $0.45\;\text{nm}$:
\begin{equation}
a_{t}=\begin{cases}
1 & \text{if }d(t)_{\text{1-13}}<0.45\ \text{nm}\\
0 & \text{otherwise}
\end{cases}\label{eq:def-contact}
\end{equation}

The three predefined states origin from the classification of the
model from Sec. \ref{par:Estimation-of-folding}. The observation
of the contact not being formed is simply $b_{t}=1-a_{t}$.

Both models, with (further called observable model) or without (ordinary
model) additional observable, take the same data splitting and the
same pretrained VAMPnet as a starting point. Afterwards both are trained
as described above except for the modified loss function with the
same time-lag values as above. The factor in front of the observable
loss is always $\lambda=10$ for the results presented here. We rotate
the observable used for training and report the results in Fig. \ref{fig:obs-stat}.
Error bars indicate the 70th percentile over 10 training runs. It
can be seen that the second timescale is confidentially overestimated
by the ordinary model (blue) as intended, which has a direct effect
on the folding and unfolding rates. Furthermore, the stationary distribution
of the three states is affected, which implies that estimates of expectation
values cannot be trusted. 

Building an observable model (orange) including the expectation of
the contact being formed improves the estimation of the stationary
distribution overall (\ref{fig:obs-stat} b). However, there is no
positive effect on the estimation of the kinetics.

Taking both autocorrelations of the contact staying formed and unformed
as observables, the observations change (green). It improves the estimates
on both stationary and kinetic properties, respectively. Whereas a
good performance on the kinetics might be expected, the reason for
the positive impact on the stationary distribution becomes only obvious
by studying the properties of the autocorrelations. The difference
between the two unnormalized autocorrelations is given by the expectation
value of the contact (Appendix \ref{par:Normalized-time-correlation}).
Therefore, if both autocorrelations are matched, the expectation value
should be matched, which has a positive effect on the state probabilities
as seen above.

Motivated by these findings we tried to match the second implied timescale
and the expectation value along training (red). Here, all observables
are matched the best. The results suggest, that as few as two experimental
observables are sufficient to counteract the bias apparent in this
data.

\section{Conclusion}

Here we extend the previously proposed reversible deep MSMs\citep{mardt2020deep}
by adding features well established for traditional MSMs: the coarse-graining
of Markov states to a fewer-state MSM, and the inclusion of experimental
restrains into the MSM estimation process. We apply these methods
to the study various aspects of the Villin headpiece miniprotein kinetics.
We exploit the fact that revDMSMs are faithful probability models
and apply transition path theory to study mean first passage times
of the folding and unfolding event, where our result coincides with
the previously published results \citep{LindorffLarsenEtAl_Science11_AntonFolding}.
Furthermore, we established an approach how experimental data can
be incorporated into the model estimation and how it can possibly
compensate for biases in the underlying force fields. The results
suggest that it is already valuable to supply few stationary and kinetic
information to the model estimation. In addition, the coarse graining
method proved valuable in constructing hierarchical models, which
give rise to easily interpretable states and allow to study the system
on different levels of detail. Finally, we demonstrated how an attention
mechanism can draw the attention to residues being important for the
classification of the dynamics. Thereby, it could be a valuable tool
for practitioners to find targets for mutations to be studied.

Despite these benefits, it remains an open challenge to develop specialized
network architectures for protein dynamics analysis, especially the
attention network could profit from an architecture where parameters
are shared among residues. Furthermore, the inclusion of real experimental
observables remains a task for future studies, where the method would
need to prove its capabilities to counteract biases of the simulation
due to the underlying force field.

\section{Acknowledgements}

This work was funded by the European Research Commission (ERC CoG
\textquotedblleft ScaleCell\textquotedblright ), Deutsche Forschungsgemeinschaft
(CRC 1114/A04, CRC 958/A04), the Berlin mathematics research center
MATH+ (Projects AA1-6 and EF1-2), and the German ministry for research
and education (BIFOLD).

\section{Data availability}

The data that support the findings of this study are available from
Lindorff-Larsen \textit{et al.}\citep{LindorffLarsenEtAl_Science11_AntonFolding}.
Restrictions apply to the availability of these data, which were used
under license for this study. Data are available from the authors
upon reasonable request and with the permission of Lindorff-Larsen
\textit{et al.}\citep{LindorffLarsenEtAl_Science11_AntonFolding}.

\bibliographystyle{unsrt}
\bibliography{all}

\section{Appendix}

\subsection{Interpretability via attention mechanism\label{subsec:Interpretability-via-attention}}

\begin{figure*}
\includegraphics[height=0.25\textheight]{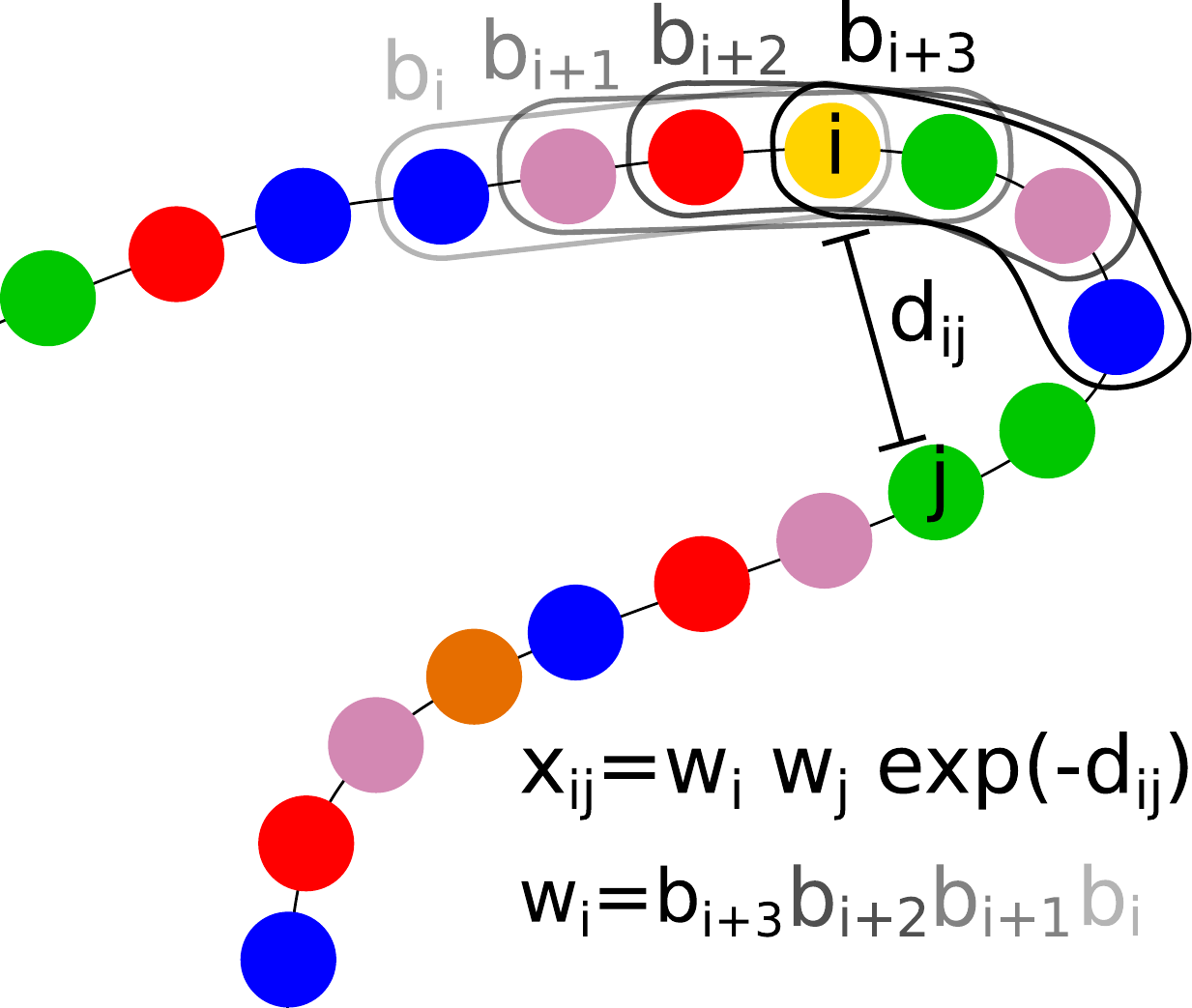}

\caption{Schematic of the attention mechanism, how to assign attention weights
for the input distances given the output values of the attention network
$\mathbf{b}(t)=\mathbf{A}(\mathbf{x}(t))$. The weight $w_{i}$ of
residue $i$ is constituted of all window size $B=4$ many values
$w_{i}=\prod_{l=0}^{B-1}b_{i+l}$. A weight $b_{i}$ is shared among
all neighbored residues within the window $i$ along the chain. Thereby,
primary structural information is fed into the network. Since a softmax
output function for the network $\mathbf{A}$ is chosen, the values
$\mathbf{b}$ can be interpreted as importance values for the classification.
By applying a softmax function to the weights $\mathbf{w}$ the same
interpretation holds for the residue weights. Since we pass $x_{ij}=\exp(-d_{ij})$
into $\boldsymbol{\chi}$, we scale each input distance $d_{ij}$
with the corresponding weights for the residues $i,j$: $x_{ij}=w_{i}w_{j}\exp(-d_{ij})$.
\label{fig:Schematic-of-the}}
\end{figure*}

In applications it is of great interest to understand how the network
defines states. For proteins this translates to understand which residues
are important for the classification. Here, we focus on an attention
formalism using an additional network $\mathbf{A}(\mathbf{x}_{t})$,
which takes a configuration $\mathbf{x}_{t}$ as an input. The input
of the classification network transforms to $\boldsymbol{\chi}(\mathbf{A}(\mathbf{x}_{t})\odot\mathbf{x}_{t})$,
where the configuration is scaled by the element wise multiplication
with the output of the attention network \citep{mnih2014recurrent,xu2015show}.
Due to a softmax output function the attention network indicates how
important each dimension of the input is. 

Since for proteins it is preferable to use internal coordinates as
inputs such as distances between residues $d_{ij}$, but scores for
each residue are easier interpretable, we define $\mathbf{A}$ to
predict weights for each residue $w_{i}$. Thereby, the input to $\boldsymbol{\chi}$
will be scaled as $x_{ij}=w_{i}w_{j}f(d_{ij})$, where $f$ can be
an arbitrary function. Additionally, we add primary information of
the chain of residues by estimating the weights $w_{i}$ via values
$\mathbf{b}=\mathbf{A}(\mathbf{x}_{t})$, which are shared within
a window along the chain (Fig. \ref{fig:Schematic-of-the}). With
the window size $B$, a weight of a residue $w_{i}$ is given by:
\begin{equation}
w_{i}=\prod_{j=0}^{B-1}b_{i+j},\label{eq:att-b}
\end{equation}

where the attention network $\mathbf{A}$ outputs a vector of dimension
$\text{dim}(\mathbf{b})=N+B-1$ assigning with a softmax function
importance values to each window. It can be interpreted as a sliding
window passing over the $N$ residue chain, which will result into
a smoother attention representation along the chain. Afterwards, the
softmax function can be applied over the weights of the residues:
\begin{equation}
\bar{w}_{i}=\frac{\exp(w_{i})}{\sum_{j}\exp w_{j}}.\label{eq:att-w}
\end{equation}

In the case of large proteins, where a lower resolution of attention
weights might be favorable, an additional skip value can be introduced
defining how many residues are skipped before a new window is defined.
The weights of the attention network are updated with a smaller learning
rate during the pretraining of the VAMPnet, since the gradients acting
from the VAMPnet $\boldsymbol{\chi}$ could be misleading due to an
unfavorable starting configuration. We are aware that the architectures
of neural network are fast developing, nonetheless we think that the
sliding window formulation is of general benefit due to its flexibility. 

\subsection{Matching two dependent time correlations\label{par:Normalized-time-correlation}}

Given a microscopic observable $a_{1}$ and defining $a_{2}=1-a_{1}$,
the expectation value of $a_{2}$ is:
\begin{equation}
\mathbb{E}[a_{2}]=\mathbb{E}[1-a_{1}]=1-\mathbb{E}[a_{1}].\label{eq:exp_value2}
\end{equation}

Therefore, it is of no use to match both expectation values $\mathbb{E}[a_{1}],\mathbb{E}[a_{2}]$.
However, if we inspect the time correlation:
\begin{align}
\mathbb{E}[a_{2}(t)a_{2}(t+\tau)] & =\mathbb{E}[(1-a_{1}(t))(1-a_{1}(t+\tau))]\label{eq:autocorr2}\\
 & =1-\mathbb{E}[a_{1}(t)]-\mathbb{E}[a_{1}(t+\tau)]+\mathbb{E}[a_{1}(t)a_{1}(t+\tau)],\nonumber 
\end{align}
it is obvious that matching both time correlations is equivalent to
matching one time correlation and the corresponding expectation value.

\subsection{Connection between timescales and folding/unfolding rates\label{par:Connection-between-timescales}}

Given experimental rates (time interval per event) for folding $r_{\text{on}}$
and unfolding $r_{\text{off}}$ it can be seen as a two state (folded
and unfolded) Markov process with a transition matrix $\mathbf{P}$.
The probabilities of jumping between these states within a time interval
$\tau$ can be calculated as the inverse rates $k_{\text{on}}=\frac{1}{r_{\text{on}}}\tau$.
The transition matrix reads then:
\begin{equation}
\mathbf{P}=\begin{bmatrix}1-k_{\text{on}} & k_{\text{on}}\\
k_{\text{off}} & 1-k_{\text{off}}
\end{bmatrix}.\label{eq:folding_rates_T}
\end{equation}

The eigenvalues of $\mathbf{P}$ are $[1,\lambda]$, where $\lambda$
is the eigenvalue corresponding to the folding process. Via the trace
the connection is recognizable:
\begin{align}
\text{tr}(\mathbf{P}) & =1+\lambda=1-k_{\text{on}}+1-k_{\text{off}}\nonumber \\
\Rightarrow\lambda & =1-k_{\text{on}}-k_{\text{off}}=1-\tau(\frac{1}{r_{\text{on}}}+\frac{1}{r_{\text{off}}}).\label{eq:folding_rates_con}
\end{align}

\subsection{Pretraining the VAMPnet\label{par:Pretraining-the-VAMPnet}}

In the original paper a pretraining of the VAMPnet with a symmetrized
VAMP-loss is recommended in order to achieve a more crisp assignment
of the classification network. Since there is no clear motivation
about this particular procedure, we modified the pretraining by adding
to the VAMP-2 score the term $\text{tr}(\mathbf{C}_{00})$ which will
maximize the eigenvalues of the matrix and therefore favors harder
state assignments. The updated loss changes to the following:
\begin{equation}
L=-\text{VAMP-2}-\lambda\text{tr}(\mathbf{C}_{00}),\label{eq:loss_updated}
\end{equation}

where $\lambda$ balances the two terms and can be set to zero during
the following unperturbed training phase. Thereby, we give a clear
motivation that the additional term pushes the network into a more
favorable region during training.
\end{document}